\documentclass[prl,twocolumn,showpacs,preprintnumbers,amsmath,amssymb]{revtex4}
\usepackage{graphicx}
\usepackage{dcolumn}
\usepackage{color}

\newcommand{\be}{\begin{equation}}
\newcommand{\ee}{\end{equation}}

\def\eg{{\it e.g.}\ } 
\def\etal{{\it et al.}} 
\def\ie{{\it i.e.}\ }

 \newcommand{\bv}{\mathbf{v}}

\begin{document}
\title{The imprint of large-scale flows on turbulence}

\author{A. Alexakis, P.D. Mininni and A. Pouquet}
\affiliation{NCAR, P.O. Box 3000, Boulder, Colorado 80307-3000, U.S.A. 
}
\date{\today}

\begin{abstract}
 
We investigate the locality of interactions in hydrodynamic turbulence using 
data from a direct numerical simulation on a grid of $1024^3$ points; the 
flow is forced with the Taylor-Green vortex. An inertial range for the energy 
is obtained in which the flux is constant and the spectrum follows an 
approximate Kolmogorov law. Nonlinear triadic interactions are dominated by 
their non-local components, involving widely separated scales. The resulting 
nonlinear transfer itself is local at each scale but the step in the energy 
cascade is independent of that scale and directly related to the integral 
scale of the flow. Interactions with large scales represent 20\% of the total energy flux.
Possible explanations for the 
deviation from self-similar models, the link between these findings and 
intermittency, and their consequences for modeling of turbulent flows 
are briefly discussed.
\end{abstract}
\maketitle

Flows in nature are often in a turbulent state driven by large scale forcing 
(\eg novae explosions in the interstellar medium) or by instabilities 
(\eg convection in the sun). Such flows involve a huge number of coupled 
modes 
leading to great complexity both in their temporal 
dynamics and in the physical structures that emerge. Many scales are excited, 
for example  from the planetary scale to the kilometer 
for convective clouds in the atmosphere, and much smaller scales  when considering 
microprocesses such as droplet formation. The question then arises 
concerning the nature of the interactions between such scales: 
are they predominantly local, involving only eddies of similar size, or 
are they as well non-local? It is usually assumed that the dominant mode 
of interaction is the former, and this hypothesis is classically viewed as 
underlying the Kolmogorov phenomenology that leads to the prediction of a 
$E(k)\sim k^{-5/3}$ energy spectrum; such a spectrum has been observed in 
a variety of contexts although there may be small corrections to this 
power-law due to the presence in the small scales of strong localized 
structures, such as vortex filaments \cite{kaneda}.

Several studies have been devoted to assess the degree of locality of 
nonlinear interactions, either through modeling of turbulent flows, as 
is the case with rapid distortion theory (RDT) \cite{LDN01} or Large Eddy 
Simulations (LES) \cite{Zhou}, or through the analysis of direct numerical 
simulations (DNS) of the Navier-Stokes equations (see \eg  
\cite{Zhou,AD90,BW94}), and more recently through rigorous bounds 
\cite{Eyink}. The spatial resolution in the numerical investigations 
was moderate, without a clearly defined inertial range and the 
differentiation between local and non-local interactions was somewhat 
limited. Thus, a renewed analysis at substantially higher Reynolds 
numbers in the absence of any modeling is in order; we address this issue 
by analyzing data stemming from a newly performed DNS on a grid of $1024^3$ 
points using periodic boundary conditions. 

The governing Navier-Stokes equation for an incompressible velocity field 
$\bv$, with ${\cal P}$ the pressure,
${\bf F}$ a forcing term and  $\nu=3\times10^{-4}$ the viscosity, reads:
\begin{equation}
\frac{\partial {\bf v}}{\partial t}+ {\bf v} \cdot \nabla {\bf v} =
-\nabla {\cal P} +\nu \nabla^2 \bv + {\bf F} \label{NS} 
\end{equation}
together with ${\bf \nabla} \cdot {\bf v} =0$. Specifically, we consider 
the swirling flow resulting from the Taylor-Green vortex~\cite{meb}:
\begin{equation} 
{{\bf  F}_{\rm TG}(k_0)}= { 2F } \,  \left[ 
\begin{array}{c} 
\sin(k_0~x) \cos(k_0~y) \cos(k_0~z) \\ 
- \cos(k_0~x) \sin(k_0~y) \cos(k_0~z)\\ 0  
\end{array} \right]  \ ,
\label{eq:Ftg}
\end{equation} 
with $k_0=2$. 
This forcing generates cells that have locally differential rotation and 
helicity, although its net helicity is zero. The resulting flow models 
the fluid between two counter-rotating cylinders \cite{meb} and has been 
used widely to study turbulence, including studies in the context of the 
generation of magnetic fields through dynamo instability 
\cite{GydroSpecialIssue}. The Reynolds number based on the integral 
scale $L \equiv 2\pi \int E(k)k^{-1} d k/E \approx 1.2 $ (where $E$ is the 
total energy), is $R_e \equiv U L/\nu \approx 4000$, where $U$ is the 
r.m.s velocity. The Reynolds number based on the Taylor scale 
$\lambda \equiv 2\pi (E/\int k^2E(k)d k)^{1/2} \approx 0.24 $, is 
$R_{\lambda} \approx 800$.

The code uses a dealiased pseudo-spectral method, with maximum wavenumber 
$k_{max}=341$ and $k_{max}\eta=1.15$, where 
$2\pi \eta=2\pi (\nu^3/\epsilon)^{1/4}$ is the dissipation scale and 
$\epsilon$ is the energy injection rate:
the flow is 
sufficiently resolved since $1/\eta$ is within the boundaries of the 
wavenumbers handled explicitly in the computation.

\begin{figure}
\centerline{\includegraphics[width=8.3cm]{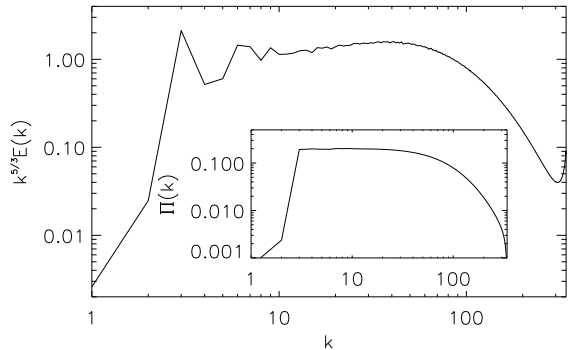}}
\caption{Compensated energy spectrum and (inset) absolute value of the 
energy flux $\Pi(k)$ in the stationary regime.}
\label{E(K)}
\end{figure}

Details of the flow dynamics will be reported elsewhere; suffice it to say 
that the flow reproduces classical features of isotropic turbulence 
\cite{Frisch_book}: the energy spectrum is well-developed (see Fig. 
\ref{E(K)}) with a constant energy flux for $k\in [5,20]$ and maximally 
helical vortex tubes are found, as predicted in 
\cite{moffatt} and shown in \cite{moffatt_tsinober,farge}. Finally, 
the anomalous exponents of longitudinal structure functions are in 
excellent agreement with previous studies \cite{kaneda} up to order 
$p=8$ (see Table I), including analysis without using the extended 
self-similarity (ESS) hypothesis \cite{ESS}.

\begin{table}[h!]
\caption{\label{tab1} Order $p$ and anomalous exponents $\zeta_p$ computed 
on two snapshots of the velocity field using the interval of scales with 
constant energy flux; the anomalous exponents $\zeta_p^{ESS}$ are computed 
using the ESS hypothesis.}
\begin{ruledtabular}
\begin{tabular}{ccccccccc}
p & 1 & 2 & 3 & 4 & 5 & 6 & 7 & 8\\ \hline
$\zeta_p$ &0.366 & 0.704 & 1.005 &1.271 & 1.502 & 1.703 & 1.878 & 2.029 \\
$\zeta_p^{ESS}$ &0.364 & 0.699 & 1 &1.264 &1.495 & 1.695 &  1.869 &  2.020
\end{tabular}
\end{ruledtabular}
\end{table} 

\begin{figure}
\centerline{\includegraphics[width=8.3cm]{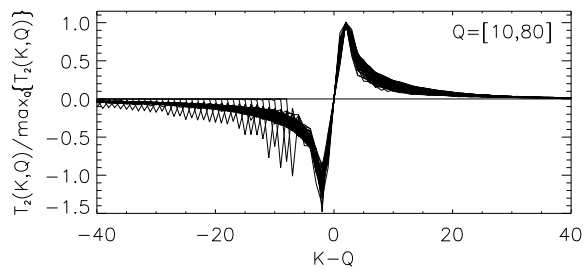}}
\caption{Normalized energy transfer from the shell Q to the shell K with
$Q \in [10,80]$. The width of the lobes is independent of K and all the 
peaks are at $K-Q \sim k_0$.}
\label{trans_fig} 
\end{figure}

To investigate the interactions between different scales we split the 
velocity field into spherical shells in Fourier space of unit width, 
\ie $\bv=\sum_K {\bf v}_K$ where ${\bf v}_K$ is the filtered velocity 
field with $K\le|k|<K+1$  (from now on called shell $K$) \cite{note1}. 
Usually, octave bands are used to define the shells (\ie shells 
of width $\Delta K^n$ are used, where $\Delta K$ is a constant often set 
to 2). This choice is based on the usual hypothesis that interactions 
are mostly local and self-similar in Fourier space, \ie the nonlinear 
term in eq. (\ref{NS}) couples triads of modes $(k,p,q)$ in Fourier 
space with $k \sim p \sim q$. Since we want to verify these hypotheses, 
we choose to use a linear step for the shells. This election does not 
imply any loss of generality, and if interactions are indeed local our 
results should be compatible with results using octave bands.

From equation (\ref{NS}), the rate of energy transfer $T_3(K,P,Q)$ (a 
third-order correlator) from energy in shell $Q$ to energy in 
shell $K$ due to the interaction with the velocity field in shell $P$ is 
defined as usual \cite{kraichnan,Alexakis05} as:
\be
T_3(K,P,Q) = -\int \bv_K \cdot (\bv_P \cdot \nabla) \bv_Q d{\bf x}^3 \ .
\label{triad_eq}
\ee
If we sum over the middle wave number $P$ we obtain the total energy 
transfer $T_2(K,Q)$ from shell $Q$ to shell $K$:
\be
T_2(K,Q) = \sum_P T_3(K,P,Q) = -\int \bv_K \cdot (\bv \cdot \nabla) 
\bv_Q d{\bf x}^3 \ .
\label{trans_eq}
\ee
Positive transfer implies that energy is transfered from shell $Q$ to $K$, 
and negative from $K$ to $Q$; thus, both $T_3$ and $T_2$ are antisymmetric 
in their $(K,Q)$ arguments. $T_2(K,Q)$ gives information on the 
shell-to-shell energy transfer between $K$ and $Q$, but not about the 
locality or non-locality of the triadic interactions themselves. 
The energy flux plotted in Fig. \ref{E(K)}is reobtained from these 
transfer functions as 
$\Pi(k) = -\sum_{K=0}^k T_1(K)  = -\sum_{K=0}^k \sum_Q T_2(K,Q)$. Note 
that the transfer terms defined in Eqs. (\ref{triad_eq},\ref{trans_eq}) 
are integrated over all volume in real space. Since in periodic boundary 
conditions there is no net flux of energy through the walls, this is 
enough to ensure that contributions due to advection (which do not lead 
to cascade of energy to smaller scales) cancel out (see \eg \cite{Eyink}).

Figure \ref{trans_fig} shows the energy transfer $T_2(K,Q)$ plotted as a 
function of $K-Q$  for 70 different values of $Q$ varying from 10 to 80. 
For each value of $Q$, the $x$ axis shows the different $K$ shells 
giving or receiving energy from that shell $Q$. All curves collapse 
to a single one: the energy in shell $K$ is received locally from shells 
with wavenumber $K-\Delta_K$ and deposited mostly in the vicinity of 
$K+\Delta_K$, with $\Delta_K\sim k_0$ for all values in the inertial range. 
In other words, the integral scale of the flow, related to the forcing 
scale $k_0^{-1}$, plays a determinant role in the process of energy 
transfer. As a result, the transfer is not self-similar, and the 
integral length scale is remembered even deep inside the constant-flux 
inertial range.

This break down of self-similarity indicates that dominant triadic 
interactions can be non-local. To examine further this point, we need 
to investigate individual triadic  interactions between Fourier 
shells by considering the tensorial transfer $T_3(K,P,Q)$. We will 
study three values of $Q$, ($Q=10$, $20$, and $40$); for each case, $P$ 
will run from 1 to 80, and $K$ from $Q-12$ to $Q+12$.

\begin{figure}
\centerline{\includegraphics[width=8.3cm]{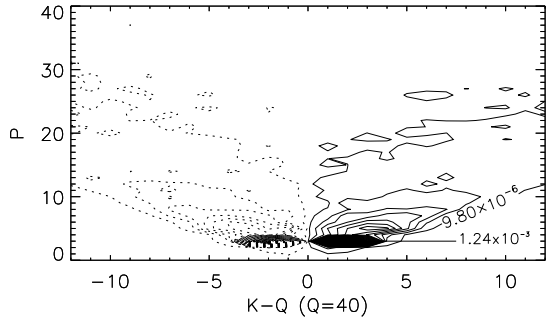}}
\caption{Contour levels of the transfer function $T_3(K,P,Q)$ for 
$Q=40$. Solid lines correspond to positive transfer, and dotted lines 
to negative transfer.} 
\label{triad_fig1}
\end{figure}

In figure \ref{triad_fig1} we show contour levels of the transfer 
$T_3(K,P,Q)$ for $Q=40$. This figure represents energy going from a 
shell $Q$ to a shell $K$ through interactions with modes in the shell 
$P$. As in Fig. \ref{trans_fig}, positive transfer means the shell $K$ 
receives energy from the shell $Q$, while negative transfer implies the 
shell $K$ gives energy to $Q$. The strongest interactions occur with $P \sim k_0$, 
and therefore the large scale flow is involved in most of the $T_2$ 
transfer of energy from small scales to smaller scales. Note that the 
individual triadic interactions with $P \sim k_0$ and $K \sim Q \pm k_0$ 
are two orders of magnitude larger than local triadic interactions.


\begin{figure}
\centerline{\includegraphics[width=8cm]{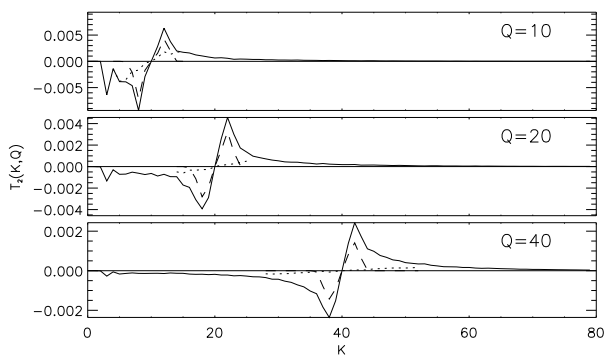}}
\caption{Comparison of the transfer functions $T_2(K,Q)$ (solid line), 
$T_2^{Loc}(K,Q)$ (dotted line), and $T_3(K,P=3,Q)$ (dashed line), 
for three values of Q.}
\label{triad_fig2}
\end{figure}

When $T_3(K,P,Q)$ in Fig. \ref{triad_fig2} is summed over all 
values of $P$, the transfer function $T_2(K,Q)$ is recovered. This 
allow us to define the transfer rate due to interactions with the 
large scale flow, and due to local interactions, summing $P$ over 
different ranges. Indeed, to further illustrate the dominance of the 
large scale flow in the involved interactions, we compare in Fig. 
\ref{triad_fig2} the total transfer function $T_2(K,Q)$ 
with the transfer due to the large scale flow $T_3(K,P=3,Q)$, and 
with the transfer due to local interactions in octave bands
$T_2^{Loc}(K,Q)=\sum_{P=Q/2}^{2Q}T_3(K,P,Q)$. The figure indicates that 
the transfer due to the local interactions ($Q/2 < P < 2Q$) is smaller 
than the transfer due to the integral length scale velocity field, and 
this behavior appears to be stronger as the value of $Q$ is increased. 
The remaining transfer comes from interactions with $P$-shells with 
wavenumbers between 1 and $Q/2$ (excluding $P=3$), which are also 
non-local in nature. Therefore, as $K$ and $Q$ get larger (as we 
go further down in the inertial range), the dominant triads 
$(K,P,Q)$ become more and more elongated, corresponding to more nonlocal 
interactions. As a result, detailed interactions between triads of modes 
are nonlocal, while the transfer of energy $T_2(K,Q)$ takes place 
between neighboring shells: local energy transfer occurs through 
non-local interactions. These results support previous claims at 
smaller resolution \cite{AD90,BW94,Zhou} that a significant 
role in the cascade of energy in the inertial range is played by 
the large scale components of the velocity field.

However, when computing the energy flux through a shell $k$, \ie 
integrating $T_2(K,Q)$ over all values of $Q$, and $K$ from $0$ to 
$k$, these non-local interactions give $\sim 20\%$ of the total flux, 
since many more local triads contribute in the global summation. We note 
that this fraction (20\%) is independent of $k$, provided that $k$ is 
large enough and in the inertial range. 


We are left therefore with two puzzles. First, why is the large scale 
flow more effective (at the level of individual triadic interactions) 
in ``destroying'' small size eddies than similar size eddies, when 
phenomenological arguments in the Kolmogorov spirit suggest otherwise? 
And secondly, why is the energy spectrum so close to $k^{-5/3}$ in the 
constant flux region, when just advection by the large scale flow would 
suggest a shallower spectrum $\sim k^{-1}$? (see \eg \cite{LDN01}). In 
what follows, we give a brief review of possible answers as well as a 
simple model that shows how a $k^{-5/3}$ energy spectrum can be obtained 
by advection and stretching of the small scales just by the large scale 
flow.

A possible answer to explain the strong non-local triadic interactions
is that the Reynolds number in the present simulation is not high enough
to observe dominance of local triads, and the decrease in amplitude of 
the small scale fields due to viscosity makes this interactions (when 
compared to the large scale flow) smaller.

Another possible answer would be that the wavenumber bands defining 
the local interactions (\ie the range of values in $P$ used to define 
$T_2^{Loc}$), that were arbitrarily taken here to have a width of $2^n$, 
could be as wide as $10^n$ as some authors suggest \cite{BW94}. If this 
is the case, a DNS with an inertial range that spans at least three 
orders of magnitude in wavenumbers would be required to actually observe 
strong local interactions. 

However, neither of these answers address the second question concerning 
why a Kolmogorov energy spectrum is observed at moderate values of 
the Reynolds number. If we look at phenomenological scaling arguments, 
we see that there is one major assumption that may not be satisfied. 
Current models assume that the energy is distributed in a hierarchy of 
vortices of size $L,L/\alpha,L/\alpha^2,...$ (with $\alpha>1$), with no 
specific geometry. However, experiments as well as numerical simulations 
have shown that enstrophy is distributed in vortex tubes, where two 
distinct length scales can be identified: one is the width of the tube 
$l$ that is typically small and varies, and one is its length $L$, 
typically of the order of the integral scale. It is not clear therefore 
when two such structures interact, which length scale is responsible for 
determining the time scale of the cascade. 

From the analysis presented here, a simple model for turbulent flows 
consistent with several features observed in simulations and 
experiments can emerge (see below). First, recall that Ref. 
\cite{farge} found that helical vortex tubes capture 99\% of the 
energy, give a $k^{-5/3}$ spectrum, and are responsible for the strong 
wings in the PDF of velocity gradients. Furthermore, it was shown 
in \cite{LDN01} that, when decomposing the velocity field in a 
large scale component $U$ and a small scale one $u$, 
artificially dropping local interactions in a simulation 
(an operation akin to RDT) gives enhanced intermittency (in the 
sense that a stronger departure from linear scaling of anomalous 
exponents is observed), while when non-local interactions are dropped 
the intermittency of the flow decreases \cite{note_MHD}.

The data analyzed in the present paper implies that, at low order of 
correlators, \ie when considering the energy flux, the interactions 
are mostly local. But when going to third-order individual triadic 
interactions (such as with $T_3$), the non-local components are dominant 
and involve the integral scale. We note that this is consistent with the 
fact that departures from a linear scaling by anomalous exponents with 
the order of structure function is stronger as the order is increased, 
since it involves more non-local interactions linked to the geometrical 
structure of vortex tubes. This leads to a model of small-scale 
interactions involving three small scales that are substantially weakened 
and gaussian, thus in agreement with the findings in \cite{LDN01} that 
such $uu$-like terms weaken intermittency as well when included in the full dynamics.

As a result, if we take into account the vortex tube structure of a 
turbulent flow, the picture of the classical Richardson cascade may 
change: a possible model to explain the aforementioned results is to 
take the time scale of the cascade as given by the geometric average 
of the length scales involved, based on the cubic root of the volume 
of the vortex tube. If this is the case, the energy dissipation 
rate of vortex tubes with velocity $u_l$ due to the large scale flow 
$U_L$ is given by $\epsilon \sim u_l^2 \cdot U_L/(l^2L)^{1/3}$. This 
implies that, for constant flux, 
$u_l \sim l^{1/3} \sqrt{\epsilon L^{1/3}/U_L}$; this scaling recovers 
the Kolmogorov spectrum, although in a different spirit \cite{note_beltrami}. 
Note that the spirit of this derivation is close to multifractal models 
used to explain intermittency corrections \cite{Frisch_book}.

Finally, we would like to point out that the success of models involving as 
an essential agent of nonlinear transfer the distortion of turbulent eddies 
by a large-scale flow -- as in RDT and its variants \cite{LDN01} or as in 
the alpha model \cite{holm} where the flow is interacting with a smooth 
velocity field (see also \cite{montgo}) -- may be in part explained by the
present findings that confirm that nonlinear triadic interactions are mostly 
nonlocal and involve the integral scale. Similar results have already been 
obtained for flows coupled to a magnetic field \cite{Alexakis05}, where the 
weakening of nonlinear interactions may occur in different fashions, \eg 
Alfv\'enization or force-free fields, and where the second order transfer 
$T_2$ between velocity and magnetic field in the induction equation is 
itself non-local.


{\it NSF grant CMG--0327888  is gratefully acknowledged. Computer time was provided by NCAR.}


\begin{thebibliography}{aa}
\bibitem{note1} With this definition, the force is acting on shell $K=3$. 
Note also we are using a sharp filter in Fourier space, and as a result 
the bounds presented in \cite{Eyink} may not apply.
\bibitem{note_beltrami}
Note that this is consistent with a Beltramization of the flow quenching 
small-scale interactions. Using the Batchelor analogy between vorticity 
and induction, and the properties of the transfer in conducting fluids
\cite{Alexakis05}, we conjecture that the vorticity transfer itself will 
be non-local;
this allows for the existence of a non-Beltrami component of nonlinear 
interactions which in turn could lead to the possible development of 
singularities.
\end{thebibliography}
\end{document}